\documentclass{article}
\usepackage{amssymb}
\usepackage{graphicx,epsfig}
\usepackage[debugshow]{tabularx}
\usepackage{ltxtable}
\usepackage{amsmath}

\begin{document}

\title{Noisy Relativistic Quantum Games in Noninertial Frames}
\author{Salman Khan$^{\dag }$\thanks{%
sksafi@phys.qau.edu.pk}, M. Khalid Khan$^{\ddag }$ \\
$^{\dag }$Department of Physics, COMSATS Institute of Information\\
Technology, Islamabad 44000, Pakistan.\\
$^{\ddag }$Department of Physics, Quaid-i-Azam University, \\
Islamabad 45320, Pakistan.}
\maketitle

\begin{abstract}
The influence of noise and of Unruh effect on quantum Prisoners' dilemma is
investigated both for entangled and unentangled initial states. The noise is
incorporated through amplitude damping channel. For unentangled initial
state, the decoherence compensates for the adverse effect of acceleration of
the frame and the effect of acceleration becomes irrelevant provided the
game is fully decohered. It is shown that the inertial player always out
scores the noninertial player by choosing defection. For maximally entangled
initially state, we show that for fully decohered case every strategy
profile results in either of the two possible equilibrium outcomes. Two of
the four possible strategy profiles become Pareto Optimal and Nash
equilibrium and no dilemma is leftover. It is shown that other equilibrium
points emerge for different region of values of decoherence parameter that
are either Pareto optimal or Pareto inefficient in the quantum strategic
spaces. It is shown that the Eisert et al \cite{Eisert} miracle move is a
special move that leads always to distinguishable results compare to other
moves. We show that the dilemma like situation is resolved in favor of one
player or the other.\newline
\end{abstract}

\section{Introduction}

Quantum game theory is a study of the well established field of classical
game theory in the light of the principals of quantum mechanics. It exploits
the remarkable properties of quantum mechanics, the entanglement and the
quantum phases, to get results that are classically impossible. For a little
more than one decade, quantum game theorists are involved to study the
behavior of classical games in the domain of quantum mechanics under
different circumstances. A number of classical games have been quantized and
the effects of entanglement on the payoff of the players have been studied.
Subsequent to the work of Meyer \cite{Meyer}, it has been shown by many
authors that quantum players can outsmart the classical counterparts by
using quantum mechanical strategies [2-13].

The entanglement between spatially separated parties is the mere powerful
source of performing various quantum information and quantum computation
processes. Its behavior in relativistic setup in noninertial frames is
presently under exploration . The existing studies in the noninertial frames
show that the Unruh effect degrades the entanglement between different modes
of various fields that may or may not vanish in the limit of infinite
acceleration \cite{Alsing,Ling,Gingrich,Pan, Schuller, Terashima}. The
studies of the effect of environment on entanglement in noninertial frames
\cite{Salman2, Jiling, Salman3} shows that the decoherence quickens the loss
of entanglement in many cases, nevertheless, under particular conditions the
entanglement rebirth may happen \cite{Salman2}.

In this paper, we study the effect of environment and of Unruh effect on the
payoffs function of the players in the quantum Prisoners' Dilemma \cite%
{Eisert}. The effect of environment is incorporated using amplitude damping
noise. We show that for a factorizable initial state and fully decohered
game, the acceleration of the frame become irrelevant and the payoff matrix
reduces to the classical symmetric payoff matrix. The local environment of
stationary observer has no effect on the payoff function of the players. For
maximally entangled initial state, the strategy profiles ($\hat{C},\hat{C}$)
and ($\hat{D},\hat{D}$) become Pareto inefficient when the game is fully
decohered. We show that in the presence of noise, the best strategy to the
classical strategy $\hat{D}$ is the quantum strategy $\hat{Q}$ (see Eq. (\ref%
{E13})). The strategy profile ($\hat{Q},\hat{C}$) is Pareto inefficient and
the strategy profile ($\hat{Q},\hat{Q}$) is Pareto optimal as well as Nash
equilibrium. We point out the origin of inconsistency between the results of
\cite{Salman4} and \cite{Eisert} and show that in the noiseless relativistic
game, the miracle move of Eisert et al \cite{Eisert} is a game winning move.
Depending on the level of noise, we show that the dilemma can be completely
or partially resolved.

\section{The Prisoners' Dilemma}

The classical Prisoners' Dilemma is a two player non-zero sum game. The
strategic space of each player consists of two strategies, cooperation ($C$)
and defection ($D$). The players (Alice and Bob) are supposed to choose a
move from their respective strategic spaces simultaneously. The reward to
the action of a player depends not only on his own move but also on the move
of his opponent. The rewards to all the possible strategic profiles of the
game are shown in the payoff matrix in Table $1$. The left number in each
pair of the matrix represents Alice's payoff and the right number in a pair
stands for Bob's payoff. This is a symmetric noncooperative game where each
player tries to maximize his/her own payoff. The catch of the dilemma is
that $D$ is the dominant strategy, that is, rational reasoning forces each
player to defect, and thereby doing substantially worse than if they would
both decide to cooperate. The behavior of the game in the domain of quantum
mechanics were first studied by Eisert et al \cite{Eisert}. The influence of
Unruh effect on the payoffs of the players in quantum prisoners' dilemma in
the relativistic setup is recently studied in Ref. \cite{Salman4}.

\begin{table*}[htb]%
\caption{Payoff matrix for the classical Prisoners' Dilemma. The first entry in a pair
of numbers denotes the payoff of Alice and the second entry represents Bob's
payoff. \label{table:1}}$%
\begin{tabular}{|c|c|c|}
\hline\hline
& Bob: $C$ & Bob: $D$ \\ \hline\hline
Alice: $C$ & $3,3$ & $0,5$ \\ \hline\hline
Alice: $D$ & $5,0$ & $1,1$ \\ \hline\hline
\end{tabular}%
$

\end{table*}%

\section{Calculation}

We consider that Alice and Bob share an entangled initial state $|\psi
_{i}\rangle _{M}=\hat{J}|0_{\omega _{A}}\rangle _{M}|0_{\omega _{B}}\rangle
_{M}$ of two fermionic qubits of mode frequencies $\omega _{A}$ and $\omega
_{B}$ at a point in flat Minkowski spacetime and all the other modes are in
vacuum state from the perspective of the inertial observer. The subscript $M$
of the kets specifies the Minkowski spacetime. The first ket is in Alice
possession and the second ket is in Bob possession. The operator $\hat{J}$\
is a symmetric unitary operator that works as an entangling gate and is
known to both players. Mathematically, it is given by%
\begin{equation}
\hat{J}=\mathrm{exp}[i\frac{\gamma }{2}\hat{D}\otimes \hat{D}],  \label{A}
\end{equation}%
where $\gamma \in \lbrack 0,\pi /2]$ and is a measure of the degree of
entanglement in the initial state. The initial state has no entanglement for
the lower limit of $\gamma $ and is maximally entangled for the upper limit
of $\gamma $. The operator $\hat{D}$ is given by%
\begin{equation}
\hat{D}=\left(
\begin{array}{cc}
0 & i \\
i & 0%
\end{array}%
\right) .
\end{equation}%
The initial state, after the entangling operator is applied, becomes%
\begin{equation}
|\psi _{i}\rangle _{M}=\cos \frac{\gamma }{2}|0_{\omega _{A}}\rangle
_{M}|0_{\omega _{B}}\rangle _{M}-i\sin \frac{\gamma }{2}|1_{\omega
_{A}}\rangle _{M}|1_{\omega _{B}}\rangle _{M}.  \label{E1}
\end{equation}%
In Eq. (\ref{E1}), $|0_{\omega _{N}}\rangle $ and $|1_{\omega _{N}}\rangle $
($N=A,B$) kets represent the vacuum and the excited states from the
perspective of an inertial observer. We consider that both the players carry
devices that are sensitive to their respective modes $\omega _{N}$. After
sharing these modes, Bob then moves with a uniform acceleration and Alice
stays stationary. The suitable coordinates for accelerated observers are
Rindler coordinates, which define two causally disconnected Rindler regions (%
$I,II$). That is, a uniformly accelerated observer in region $I$ has no
access to information that leaks to region $II$ and vice versa (for detail
see \cite{Alsing} and reference therein). A given Minkowski mode of a
particular frequency spreads over all positive Rindler frequencies that
peaks about the Minkowski frequency \cite{Takagi,Alsing2}. However, to
simplify our problem we consider a single mode only in the Rindler region $I$%
, which is valid if the observers' detectors are highly monochromatic that
detects the frequency $\omega _{A}\sim \omega _{B}=\omega $. From this point
onward, with this approximation, the frequency subscript of kets will be
dropped.

From the perspective of accelerated frame, the Minkowski vacuum state is a
two mode squeezed state given by \cite{Alsing, Aspache, Martinz, Bruche}%
\begin{equation}
|0\rangle _{M}=\cos r|0\rangle _{I}|0\rangle _{II}+\sin r|1\rangle
_{I}|1\rangle _{II},  \label{E2}
\end{equation}%
and the Minkowski excited state is given by%
\begin{equation}
|1\rangle _{M}=|1\rangle _{I}|0\rangle _{II},  \label{E3}
\end{equation}%
where $I$ and $II$ in the subscript of the kets represents the modes in the
two Rindler regions. Eq. (\ref{E2}) shows that the noninertial observer that
moves with a constant acceleration in region $I$ sees a thermal state
instead of the vacuum state. This effect is called the Unruh effect \cite%
{Davies,Unruh}. The parameter $r$ is the dimensionless acceleration
parameter given by $\cos r=\left( e^{-2\pi \omega c/a}+1\right) ^{-1/2}$.
The constants $\omega $, $c$ and $a$, in the exponential stand,
respectively, for Dirac particle's frequency, speed of light in vacuum and
Bob's acceleration. The parameter $r=0$ when acceleration $a=0$ and $r=\pi
/4 $ when $a=\infty $. Using Eqs. (\ref{E2}) and (\ref{E3}) in Eq. (\ref{E1}%
), the initial state in terms of Minkowski mode for Alice and Rindler modes
for Bob becomes%
\begin{eqnarray}
|\psi \rangle _{M,I,II} &=&\cos \frac{\gamma }{2}\cos r|0\rangle
_{M}|0\rangle _{I}|0\rangle _{II}  \notag \\
&&+\cos \frac{\gamma }{2}\sin r|0\rangle _{M}|1\rangle _{I}|1\rangle
_{II}-i\sin \frac{\gamma }{2}|1\rangle _{M}|1\rangle _{I}|0\rangle _{II}.
\label{E4}
\end{eqnarray}%
Since Bob has no access to the information in region $II$, therefore,
tracing over the modes in region $II$ gives the following mixed density
matrix%
\begin{eqnarray}
\rho _{M,I} &=&\cos ^{2}\frac{\gamma }{2}\cos ^{2}r|00\rangle \langle 00|-%
\frac{1}{2}i\sin \gamma \cos r|11\rangle \langle 00|  \notag \\
&&+\cos ^{2}\frac{\gamma }{2}\sin ^{2}r|01\rangle \langle 01|+\frac{1}{2}%
i\sin \gamma \cos r|00\rangle \langle 11|  \notag \\
&&+\sin ^{2}\frac{\gamma }{2}|11\rangle \langle 11|.  \label{E5}
\end{eqnarray}%
Note that we have dropped the subscript of the kets.

\subsection{The game in a noisy environment}

Now, we consider that prior to the execution of moves by the players, the
state of the game evolves through a noisy environment. The interaction of a
quantum system with a noisy environment can be described in terms of Kraus
operators. The final density matrix of a system when it evolves in a noisy
environment can be written as%
\begin{equation}
\rho _{f}=\sum_{i,j,k,...}K_{i}K_{j}K_{k}...\rho ...K_{k}^{\dag }K_{j}^{\dag
}K_{i}^{\dag },  \label{E6}
\end{equation}%
where $\rho $ is the initial density matrix of a system and $K_{n}$ are the
Kraus operators that describe the interaction of the system with the
environment and satisfy the completeness relation $\sum_{n}K_{n}^{\dag
}K_{n}=I$. There could be different kinds of interactions between the system
and environment and each kind of interaction has its own form of Kraus
operators. The kind of interaction that we consider here is known as
amplitude damping. The Kraus operators for amplitude damping channel of a
single qubit system are given as%
\begin{equation}
K_{o}=\left(
\begin{array}{cc}
1 & 0 \\
0 & \sqrt{1-p}%
\end{array}%
\right) ,\qquad K_{1}=\left(
\begin{array}{cc}
0 & 0 \\
0 & \sqrt{p}%
\end{array}%
\right) ,  \label{E7}
\end{equation}%
where $p$ ($0\leq p\leq 1$) is called decoherence parameter. The system is
undecohered for the lower limit and is fully decohered for the upper limit
of $p$. In the following we consider two kinds of interactions. In one case,
we study the coupling of only Bob's qubit with the environment and in the
second case both the qubits are under the influence of the environment.

The Kraus operators $E_{i}$\ for the case when only Bob's qubit is locally
under the influence of environment can be expressed in terms of the single
qubit Kraus operators as $E_{i}^{B}=I\otimes K_{i}$, where $I$ is a single
qubit identity matrix and $i=0,1$. Similarly, when Alice's qubit is locally
influenced by the environment, the Kraus operators for that case become $%
E_{i}^{A}=K_{i}\otimes I$. The density matrix of the system, after the
coupling with environment, becomes%
\begin{equation}
\rho _{M,I,E}=\sum_{i}E_{i}^{A}E_{i}^{B}\rho _{M,I}E_{i}^{B\dag
}E_{i}^{A\dag },  \label{E8}
\end{equation}%
where $E$ in the subscript of $\rho $ is a reminiscence that the game state
is influenced by the environment. The decoherence parameter for Alice's
qubit is represented by $p_{1}$ and for Bob's qubit it is given by $p_{2}$.

In the quantum Prisoners' Dilemma, the strategic moves of the players are
unitary operators that can be expressed as \cite{Flitney4}%
\begin{equation}
\hat{U}_{N}(\alpha ,\theta )=\left(
\begin{array}{cc}
e^{i\alpha _{N}}\cos \frac{\theta _{N}}{2} & i\sin \frac{\theta _{N}}{2} \\
i\sin \frac{\theta _{N}}{2} & e^{-i\alpha _{N}}\cos \frac{\theta _{N}}{2}%
\end{array}%
\right) ,  \label{E9}
\end{equation}%
where the subscript $N=A,B$ stands for Alice and Bob, $\theta \in \lbrack
0,\pi ]$ and $\alpha \in \lbrack -\pi ,\pi ]$. More generally, the quantum
mechanical strategic spaces of the players can be expressed in terms of
unitary operators made up of three parameters \cite{Benjamin,Flitney4}.
Nevertheless, in the present work we will use the two parameter strategic
spaces of Eq. (\ref{E9}). If the move cooperation of the players is
associated with state $|0\rangle $ and the move defection is associated with
state $|1\rangle $, then the quantum strategy $\hat{C}$ corresponds to $\hat{%
U}_{N}(0,0)$ and the quantum strategy $\hat{D}$ corresponds to $\hat{U}%
_{N}(0,\pi )$. After executing their moves, the final density matrix of the
game prior to the measurement becomes \cite{Eisert}%
\begin{equation}
\rho =\hat{J}^{\dag }\left( \hat{U}_{A}\otimes \hat{U}_{B}\right) \rho
_{M,I,E}\left( \hat{U}_{A}^{\dag }\otimes \hat{U}_{B}^{\dag }\right) \hat{J},
\label{E10}
\end{equation}%
where the disentangling gate $\hat{J}^{\dag }$\ is applied to disentangle
the final density matrix in order to make measurement of the payoffs. The
expected payoffs of the players are then found by using the following
equation%
\begin{equation}
P_{N}^{j_{1}j_{2}}=\sum_{i}\$_{N}^{j_{1(i)}j_{2(i)}}\rho _{ii},  \label{E11}
\end{equation}%
where $\rho _{ii}$ ($i\in \lbrack 0,1]$) are the diagonal elements of the
final density matrix and $\$_{N}^{j_{1}(i)j_{2}(i)}$ ($j_{1(i)},j_{2(i)}\in
\lbrack C,D]$) are the classical payoffs of the players from table $1$
whereas $j_{1},j_{2}$ in the superscript of $P$ could be any of the
classical or quantum strategies.

\section{Results and Discussion}

The payoffs for initially unentangled state ($\gamma =0$) when the players
are restricted to the classical strategic spaces are given in table $2$. In
each cell of the table, the top payoff corresponds to Alice and the bottom
one corresponds to Bob. It can easily be seen from the table that the
results of Ref. \cite{Salman4} are retrieved for $p_{2}=0$ and for $%
p_{2}=r=0 $, the classical results are obtained. Generally, the symmetry of
payoff matrix that exists in the classical form of the game is lost and non
of the strategy profiles results in same payoffs to both players. However,
regarding decoherence, there are two other important features of the payoff
matrix of the table that need to be pointed out, (a) the payoffs depend only
on the decoherence parameter $p_{2}$, which means that for unentangled
initial state the coupling of the stationary player with the environment
have no effect on the payoffs of the players. This is an important result,
as for practical applications of the game, one doesn't need to bother about
the local coupling of the inertial player with the environment. (b) each
payoff changes by an amount $\pm np_{2}\sin ^{2}r$, where $n$ is an integer,
such that for a fully decohered ($p_{2}=1$) case the acceleration parameter $%
r$ becomes irrelevant and the symmetric classical payoff matrix is
retrieved. For a fully decohered case, the decoherence compensates for the
asymmetry of the payoff matrix caused by the acceleration of the frame and
brings fairness back to the game. Although,for any sets of values of $p_{2}$
and $r$, Alice dominates by playing $\hat{D}$, however, the dilemma of the
classical game exists as the strategy profiles ($\hat{C},\hat{C}$) and ($%
\hat{D},\hat{D}$) are respectively Pareto optimal and Nash equilibrium.

\begin{table*}[htb]%
\caption{The payoff matrix of the players' payoffs as a function of the
acceleration of Bob's frame and decoherence parameter. The top  entry in every cell corresponds to
Alice's payoff and the bottom entry corresponds to Bob's payoff. The initial state of the game
is unentangled and the players are allowed to select a move from the two
pure classical moves. \label{table:1}}$%
\begin{tabular}{|c|c|c|}
\hline
& Bob: $\hat{C}$ & Bob: $\hat{D}$ \\ \hline
Alice: $\hat{C}$ & $3\left( \cos ^{2}r+p_{2}\sin ^{2}r\right) ,$ & $3\left(
1-p_{2}\right) \sin ^{2}r,$ \\
& $4-\cos 2r-2p_{2}\sin ^{2}r$ & $4+\cos 2r+2p_{2}\sin ^{2}r$ \\ \hline
Alice: $\hat{D}$ & $3+2\cos 2r+4p_{2}\sin ^{2}r,$ & $3-2\cos 2r-4p_{2}\sin
^{2}r,$ \\
& $\left( 1-p_{2}\right) \sin ^{2}r$ & $\cos ^{2}r+p_{2}\sin ^{2}r$ \\ \hline
\end{tabular}%
$

\end{table*}%
\begin{figure}[!h]
\centering
\begin{tabular}{cc}
\epsfig{file=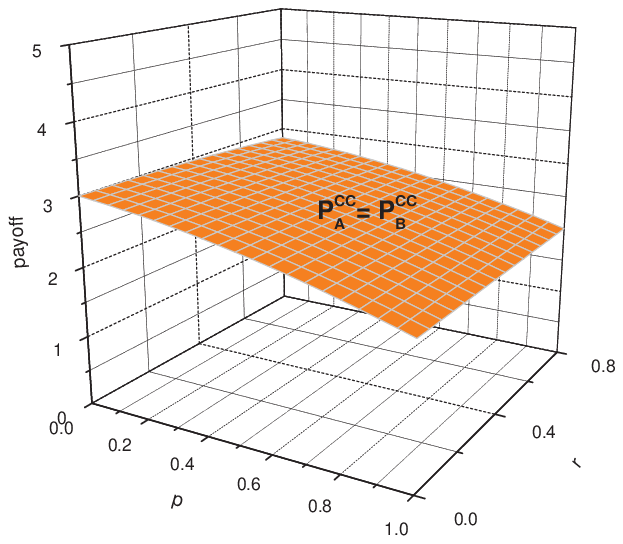,width=0.5\linewidth,clip=} & %
\epsfig{file=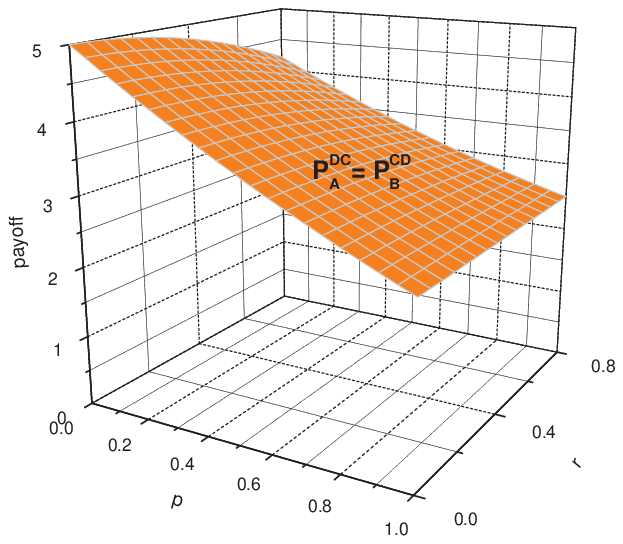,width=0.5\linewidth,clip=} \\
\epsfig{file=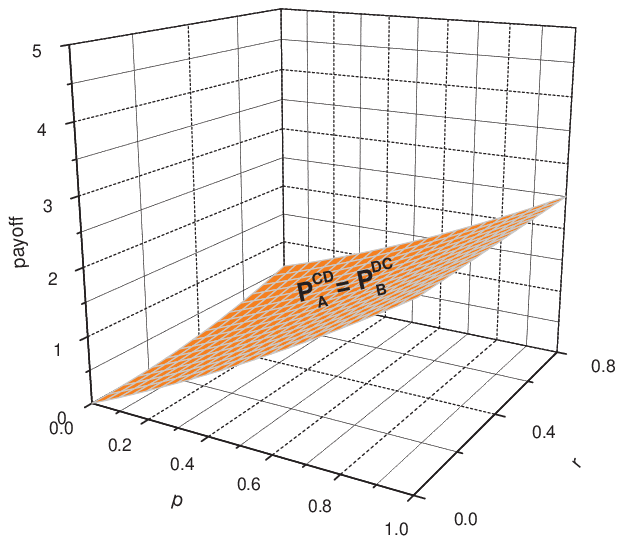,width=0.5\linewidth,clip=} & %
\epsfig{file=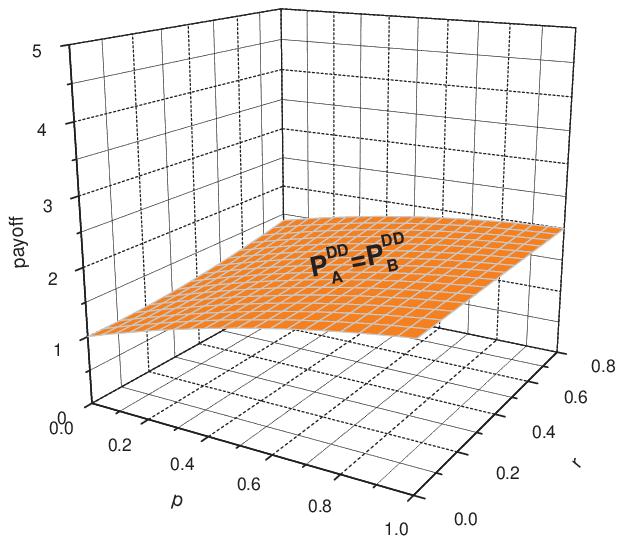,width=0.5\linewidth,clip=}%
\end{tabular}%
\caption{(color online) The payoffs for different strategy profiles are
plotted against the decoherence parameter $p$ and the acceleration parameter
$r$ for the case when the initially state is maximally entangled.}
\label{Figure1}
\end{figure}

\begin{figure}[h]
\begin{center}
\begin{tabular}{ccc}
\vspace{-0.5cm} \includegraphics[scale=1.2]{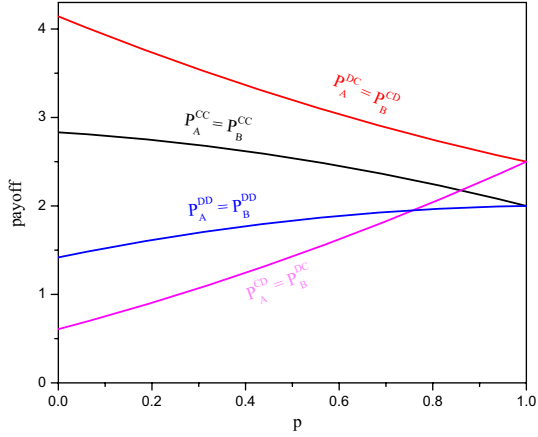}\put(-320,220) &  &
\end{tabular}%
\end{center}
\caption{(color online) The payoffs for different strategy profiles are
plotted against the decoherence parameter $p$ for the case when the
initially state is maximally entangled and the acceleration parameter $r=%
\protect\pi /4$.}
\label{Figure2}
\end{figure}

Now, we investigate the effects of decoherence on payoff functions of the
players for maximally entangled initial state ($\gamma =\pi /2$) by
restricting the players first to the classical strategic spaces and then
allowing the players to choose their moves from quantum mechanical strategic
spaces as well. For both of these cases, the payoff matrices are very
interesting and the results are drastically different from the payoff matrix
of the unentangled initial state. We begin for the case when both players
are bound to the classical strategic spaces. The payoffs for all the four
possible strategy profiles become%
\begin{eqnarray}
P_{A,B}^{CC(DD)} &=&\frac{17}{8}+\frac{p_{1}}{4}\pm \sqrt{(1-p_{1})(1-p_{2})}%
]\cos r-\frac{1}{8}\cos 2r  \notag \\
&&+p_{2}(\frac{1}{8}-\frac{p_{1}}{2}+\frac{1}{8}\cos 2r),  \notag \\
P_{A}^{CD(DC)} &=&P_{B}^{DC(CD)}=\frac{19}{8}-\frac{p_{1}}{4}\mp \frac{5}{2}%
\sqrt{(1-p_{1})(1-p_{2})}]\cos r+\frac{1}{8}\cos 2r  \notag \\
&&-p_{2}(\frac{1}{8}-\frac{p_{1}}{2}+\frac{1}{8}\cos 2r),  \label{E12}
\end{eqnarray}%
where the upper sign in$"\pm "$ and in $"\mp "$ corresponds to the first
strategy profile and the lower sign corresponds to the second strategy
profile, bracketed in the superscript of each payoff function. For strategy
profiles ($\hat{C},\hat{C}$) and ($\hat{D},\hat{D}$), the results of Refs.%
\cite{Eisert,Salman4} are retrieved by setting, respectively, the
corresponding parameters equal to zero. For strategy profile ($\hat{C},\hat{D%
}$) and ($\hat{D},\hat{C}$), setting $p_{i^{\prime }}s=0$, the results are
inverted to the results obtained in \cite{Salman4}, however, these are in
agreement with the results of Ref. \cite{Eisert} for $p_{i^{\prime }}s=r=0$.
The inversion of payoffs between the players for strategy profiles ($\hat{C},%
\hat{D}$) and ($\hat{D},\hat{C}$) in \cite{Salman4} happens because of the
use of non-symetric form \cite{Benjamin,Flitney4} of operator $\hat{D}_{1}$
(Eq. (2)) for constructing the entangling operator $\hat{J}$ . The choice of
operator $\hat{D}$, in the present work, for constructing the entangling
operator $\hat{J}$ removes this mismatch. Unlike the payoff matrix for
unentangled initial state (table $2$), the payoff matrix of Eq. (\ref{E12})
is a symmetric payoff matrix and each payoff depends on coupling with the
two local environments. Furthermore, the strategy profiles ($\hat{C},\hat{C}$%
) and ($\hat{D},\hat{D}$) define two different equilibrium outcomes, like
the undecohered case of the game in noninertial frames. Since the payoffs
become functions of decoherence parameters $p_{i^{\prime }}s$, we want to
see how the payoffs vary with these parameters by considering the simplest
case with $p_{1}=p_{2}=p$. This means that instead of two different local
environments, both qubits are coupled together to a single collective
environment. The payoffs for different strategy profiles are plotted against
the decoherence parameter $p$ and the acceleration parameter $r$ in figure %
\ref{Figure1}. One can see from the figure that $\hat{D}$ is the dominant
strategy. Again, this result is not in agreement with the result of \cite%
{Salman4} for the same reason as mentioned above. The effect of decoherence
is not identical for every strategy profile rather it's somewhat additive to
the effect of acceleration parameter $r$. For example, in the first row of
figure \ref{Figure1}, both $p$ and $r$ reduce the payoffs whereas in the
second row both act to increase the payoffs keeping one or the other
parameter constant. Also, it is clear from the figure that the strategy
profile ($\hat{C},\hat{C}$) becomes Pareto Optimal and the strategy profile (%
$\hat{D},\hat{D}$) becomes Nash equilibrium. This is true for the whole
range of decoherence parameter $p$ in the region of lower values of the
acceleration of the accelerated frame. In the range of infinite acceleration
and upper values of $p$ the behavior of the game changes. This change in the
behavior of the game is shown in figure \ref{Figure2}, where payoffs of Eq. (%
\ref{E12}), corresponding to different strategy profiles, are plotted
against decoherence parameter for $r=\pi /4$. It can be seen from the figure
that, although, $\hat{D}$ is still the dominant strategy, however, for $%
0.85<p<1$, the payoffs corresponding to the two equilibrium strategy
profiles (($\hat{C},\hat{C}$) and ($\hat{D},\hat{D}$)) are no good for any
player. The most striking feature of figure \ref{Figure2} corresponds to the
fully decohered game ($p=1$) where every strategy profile results into one
of the two outcomes. One can see that the strategy profiles ($\hat{C},\hat{C}
$) and ($\hat{D},\hat{D}$) become \textit{Pareto inefficient} whereas the
strategy profiles ($\hat{C},\hat{D}$) and ($\hat{D},\hat{C}$) are both
Pareto Optimal and Nash equilibrium and there is no dilemma left in the
game. Regardless of the move of a player, the game ends up in either of the
two outcomes.

Now, we turn to the case where players have access to the quantum mechanical
strategic spaces. At this end, we first study the effects of decoherence on
the payoff function of the players when one or both of them plays the Eisert
\textit{et al.} \cite{Eisert} quantum mechanical strategy $\hat{Q}$ given by%
\begin{equation}
\hat{Q}=\hat{U}\left( \pi /2,0\right) =\left(
\begin{array}{cc}
i & 0 \\
0 & -i%
\end{array}%
\right) .  \label{E13}
\end{equation}%
Let Alice play $\hat{Q}$, then, the payoffs as function of the remaining
four parameters ($p,r,\alpha _{B},\theta _{B}$) are given by%
\begin{eqnarray}
P_{A,B}^{Q\theta _{B}} &=&\frac{1}{8}[18-\{1-(3-4p)p+(1-p)\cos 2r\}\cos
\theta _{B}  \notag \\
&&-2(1-p)\cos r\{2\cos 2\alpha _{B}(1+\cos \theta _{B})\pm 5\cos \theta
_{B}\mp 5\}].  \label{E14}
\end{eqnarray}%
In Eq. (\ref{E14}), the upper sign in $"\pm "$ and in $"\mp "$ stands for
Alice's payoff and the lower one for Bob's payoff. The results of Eq. (\ref%
{E14}) are, again, inverted to the results of Ref. \cite{Salman4} for $p=0$
and reduces to the result of \cite{Eisert} for setting both $r$ and $p$
equal to zero. Also, a close look on the equation immediately reveals that
for a fully decohered game ($p=1$), the effects of acceleration $r$ and
quantum phase $\alpha _{B}$ become irrelevant. Under such condition $%
P_{A,B}^{Q\theta _{B}}=1/8(18-2\cos \theta _{B})$ that results in large
payoffs when Bob executes $\hat{D}$. Similarly, for $p<1$, the strategy
profiles ($\hat{Q},\hat{C}$) and ($\hat{Q},\hat{Q}$) are two equilibrium
outcomes (at which $P_{A}^{QC}=P_{B}^{QC}$ and $P_{A}^{QQ}=P_{B}^{QQ}$).
However, for strategy profile ($\hat{Q},\hat{C}$) the payoffs increases with
increasing value of $p$ whereas for strategy profile ($\hat{Q},\hat{Q}$) it
decreases. As stated above, against Bob's strategy $\hat{C}$ Alice's best
strategy is $\hat{D}$ and against Bob's strategy $\hat{D}$ Alice's best
strategy is $\hat{Q}$. This means that no single strategy of Alice dominate
her against every strategy of Bob. The payoff matrix is symmetric from the
perspective of strategy $\hat{Q}$. The strategy profile ($\hat{Q},\hat{C}$)
(and hence ($\hat{C},\hat{Q}$)) is Pareto inefficient and the strategy
profile ($\hat{Q},\hat{Q}$) is both Pareto optimal and Nash equilibrium.

Finally, we investigate the effects of decoherence and acceleration on the
payoffs when one player is limited to the classical strategic space and the
other is allowed to choose any move from the quantum mechanical strategic
space. This choice of the strategic spaces make the game unfair both in
inertial and noninertial setups. In inertial setup, it is shown \cite{Eisert}
that there exists a move for the quantum player to outsmart the classical
player regardless of what the classical player is choosing from his
strategic space. This move is known, in the literature on quantum games, as
the \textit{miracle move} and is given by%
\begin{equation}
\hat{M}=\hat{U}\left( -\frac{\pi }{2},\frac{\pi }{2}\right) =\frac{i}{\sqrt{2%
}}\left(
\begin{array}{cc}
-1 & 1 \\
1 & 1%
\end{array}%
\right) .  \label{E15}
\end{equation}%
In noninertial frame, it is shown in \cite{Salman4} that playing this move
always results in a poor result for the quantum player. This result is
contradictory to the quantum mechanical results obtained in inertial setup
of the game. The disagreement between the results of the game in the two
setups, as discussed earlier, is because of the choice of choosing the
operator $\hat{D}$ for constructing the entangling operator $\hat{J}$.
\begin{figure}[h]
\begin{center}
\begin{tabular}{ccc}
\vspace{-0.5cm} \includegraphics[scale=1.2]{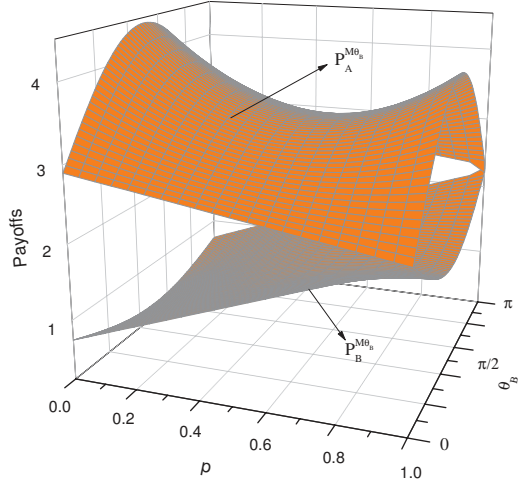}\put(-320,220) &  &
\end{tabular}%
\end{center}
\caption{(color online) The payoffs of the players are plotted against the
decoherence parameter $p$ and $\protect\theta _{B}$ when Alice plays the
miracle move $\hat{M}$ . The acceleration parameter $r=\protect\pi /6$.}
\label{Figure3}
\end{figure}

To study the effects of miracle move $\hat{M}$ in the present setup of the
game, we restrict Bob to classical strategic space and allow Alice to
execute $\hat{M}$. The payoffs under this condition become%
\begin{eqnarray}
P_{A}^{M\theta _{B}} &=&\frac{1}{4}[\frac{7}{2}\{1+p\left( 4p-3\right)
+\left( 1-p\right) \cos 2r\}\sin \theta _{B}  \notag \\
&&+\left( 1-p\right) \cos r\left( \sin \theta _{B}+3\right) +9],  \notag \\
P_{B}^{M\theta _{B}} &=&\frac{1}{4}[-\frac{3}{2}\{1+p\left( 4p-3\right)
+\left( 1-p\right) \cos 2r\}\sin \theta _{B}  \notag \\
&&+\left( 1-p\right) \cos r\left( \sin \theta _{B}-7\right) +9].  \label{E16}
\end{eqnarray}%
It can be seen from Eq. (\ref{E16}) that setting $p=0$ does not produce the
result of \cite{Salman4}, rather the results are inverted. However, setting $%
p=r=0$ produce the results of \cite{Eisert}. The payoff matrix of Eq. (\ref%
{E16}) is symmetric with respect to the interchange of the players. It can
be checked that for noiseless game, $P_{A}^{M\theta _{B}}>P_{B}^{M\theta
_{B}}$ for Bob's classical strategic space. That is, the quantum player
dominates the classical one in the noiseless relativistic setup of the game
by playing $\hat{M}$. In fact, as can be seen from Eq. (\ref{E16}), the
choice from the two classical moves is meaningless against $\hat{M}$ both
for noisy and noiseless setup of the game. However, there is a remarkable
difference between the noisy and noiseless game. That is, for a fully
decohered game ($p=1$) $P_{A}^{M\theta _{B}}=P_{B}^{M\theta _{B}}=2.25$,
irrespective of the choice of Bob ($\theta _{B}=0$ or $\pi $) and of
acceleration of the frame. Thus for a fully decohered game the unfair game
just become a fair game that always ends up in a single outcome to both
players, provided one player is playing the miracle move and the other is
limited to the classical two strategy space. Bob cannot outscore Alice even
if he has access to the entire classical space spanned by $\theta _{B}$ for
the whole range of decoherence parameter $p$ and acceleration parameter $r$.
Figure 3 shows that how the payoffs of the players varies against $p$ and $%
\theta _{B}$. One can see that against Alice strategy $\hat{M}$ the best
that Bob can gain is to play either $\hat{C}$ or $\hat{D}$. These results
show that the miracle move is a special move in every setup of the game. In
noiseless inertial and noninertial setups, it always let the player win.
Against the two classical strategies, for a fully decohered case it makes
the game fair and equally benefits the two players.

\section{Summary}

The effect of decoherence by using amplitude damping channel on the payoffs
function of the players in the quantum prisoners' dilemma in noninertial
frame has been investigated. It is shown that for unentangled initial
state,generally, no strategy profile results in equal payoffs to both
players and the game becomes an asymmetric game. Interestingly enough, the
symmetric classical payoff matrix is recovered regardless of the
acceleration of the frame for a fully decohered game. The local environment
of the inertial player has no effect on the payoff functions of the players.
The biasing effect of the acceleration of the noninertial frame is
compensated by the amplitude damping noise. The inertial player always
outscores the noninertial player by playing $\hat{D}$.

For maximally entangled initial state, it is shown that when the players are
bound to the classical strategic spaces, the payoff matrix is a symmetric
payoff matrix and the strategy profiles ($\hat{C},\hat{C}$) and ($\hat{D},%
\hat{D}$) become \textit{Pareto inefficient} and there is no dilemma left in
the game provided the game is fully decohered. For other values of
decoherence parameter $p$, the strategy profile ($\hat{C},\hat{C}$) is
Pareto Optimal and ($\hat{D},\hat{D}$) is Nash equilibrium which makes the
strategy $\hat{D}$ a dominant strategy. The payoff matrix is symmetrical
with respect to quantum strategy $\hat{Q}$ and the dominant strategy against
$\hat{D}$ is the quantum strategy $\hat{Q}$. The strategy profile ($\hat{Q},%
\hat{Q}$) is both Pareto optimal and Nash equilibrium. It is shown that the
two classical moves become irrelevant against the miracle move $\hat{M}$,
that is, both $\hat{C}$ and $\hat{D}$ lead to the same poor payoff for
classical player. However against $\hat{C}$ and $\hat{D}$, for a fully
decohered game, the miracle move reduce the game to a single outcome where
both players are equally benefited and thus throw out the unfairness that
exists in both inertial and noninertial setup of the quantum form of the
game. The miracle move proves itself to be treated as a special move under
any setup of the game.\newline


\begin{thebibliography}{99}
\bibitem{Meyer} Meyer, D. A.: Quantum Strategies. Phys. Rev. Lett. 82
1052-1055 (1999)

\bibitem{Eisert} Eisert, J., Wilkens, J., Lewenstein, M.: Quantum Games and
Quantum Strategies. Phys. Rev. Lett. 83, 3077--3080 (1999)

\bibitem{Marinatto} Marinatto, L., Weber, T.: A quantum approach to static
games of complete information. Phys. Lett. A 272, 291--303 (2000)

\bibitem{H. Li} Li, H., Du, J., Massar, S.: Continuous-Variable Quantum
Games. Phys. Lett. A 306 73-78 (2002)

\bibitem{CF Lo} Lo, C. F., Kiang, D.: Quantum Bertrand duopoly with
differentiated products. Phys. Lett. A 321 94-98 (2004)

\bibitem{Flitney1} Flitney, A. P., Abbott D.:Quantum version of the Monty
Hall problem. Phys. Rev. A 65, 062318 (2002)

\bibitem{Iqbal} Iqbal, A., Toor, A. H.: Backwards-induction outcome in a
quantum game. Phys. Rev. A 65, 052328 (2002)

\bibitem{Flitney2} Flitney, A. P., Ng, J., Abbott, D.: Quantum Parrondo's
games. Physica A 314 35-42 (2002)

\bibitem{Goldenberg} Goldenberg, L., Vaidman, L., Wiesner, S.: Quantum
Gambling. Phys. Rev. Lett. 82, 3356--3359 (1999)

\bibitem{Salman1} Khan, S., Ramzan, M., Khan, M. K.: Quantum Model of
Bertrand Duopoly. Chin. Phys. Lett. 27, 080302 (2010)

\bibitem{Chen} Chen, L. K., Ang, H., Kiang, D., Kwek. L.C. Lo, C. F.:
Quantum prisoner dilemma under decoherence. Phys. Lett. A 316 317-323 (2003)

\bibitem{Flitney3} Flitney, A. P., Abbott, D.: Quantum games with
decoherence. J. Phys. A: Math. Gen. 38 449-459 (2005)

\bibitem{Salman} Khan, S., Ramzan, M., Khan, M. K.: Quantum Parrondo's Games
Under Decoherence Int. J. Theo. Phys. 49 31-41 (2010)

\bibitem{Alsing} Alsing, P. M., Fuentes-Schuller, I., Mann, R. B., Tessier,
T. E.: Entanglement of Dirac fields in noninertial frames. Phys. Rev. A 74,
032326 (2006)

\bibitem{Ling} Ling, Y., He, S., Qiu, W., Zhang, H.: Quantum entanglement of
electromagnetic field in non-inertial reference frames.J. Phys. A: Math.
Theor. 40 9025--9032 (2007)

\bibitem{Gingrich} Gingrich, R. M., Adami, C.: Quantum Entanglement of
Moving Bodies. Phys. Rev. Lett. 89, 270402 (2002)

\bibitem{Pan} Pan, Q., Jing, J.: Degradation of nonmaximal entanglement of
scalar and Dirac fields in noninertial frames. Phys. Rev. A 77, 024302 (2008)

\bibitem{Schuller} Fuentes-Schuller, I., Mann, R. B.: Alice Falls into a
Black Hole: Entanglement in Noninertial Frames. Phys. Rev. Lett. 95, 120404
(2005)

\bibitem{Terashima} Terashima, H., Ueda, M.: Relativistic
Einstein-Podolsky-Rosen correlation and Bell's inequality. Int. J. Quantum
Inf. 1, 93-114 (2003)

\bibitem{Salman2} Khan, S., Khan, M. K.: Open quantum systems in noninertial
frames. J. Phys. A: Math. Theor. 44 045305 (2011)

\bibitem{Jiling} wang J and Jing J.: Quantum decoherence in noninertial
frames. Phys. Rev. A 82, 032324 (2010)

\bibitem{Salman3} Khan, S.: Entanglement of tripartite states with
decoherence in non-inertial frames. J. Mod. Opt. 59 250-258 (2012)

\bibitem{Salman4} Khan, S., Khan, M. K.: Relativistic quantum games in
noninertial frames. J. Phys. A: Math. Theor. 44 355302 (2011)

\bibitem{Takagi} Takagi, S.: Vacuum Noise and Stress Induced by Uniform
Acceleration. Prog. Theor. Phys. Suppl. 88 1-142 (1986)

\bibitem{Alsing2} Alsing, P. M., McMahon, D. Milburn, G. J.: Teleportation
in a non-inertial frame.  J. Opt. B: Quantum Semiclass. Opt. 6 S834--S843
(2004)

\bibitem{Aspache} Aspachs, M., Adesso, G., Fuentes, I.: Optimal Quantum
Estimation of the Unruh-Hawking Effect. Phys. Rev. Lett. 105, 151301 (2010)

\bibitem{Martinz} Martin-Martinez E, Garay L J, and Leon J.: Quantum
entanglement produced in the formation of a black hole. Phys. Rev. D 82,
064028 (2010).

\bibitem{Bruche} Bruschi, D. E., Louko, J., Martin-Martinez, E., Dragan, A.,
Fuentes, I.: Unruh effect in quantum information beyond the single-mode
approximation. Phys. Rev. A 82, 042332 (2010)

\bibitem{Davies} Davies, P. C. W.: Scalar production in Schwarzschild and
Rindler metrics. J. Phys. A 8 609-616 (1975)

\bibitem{Unruh} Unruh, W. G.: Notes on black-hole evaporation. Phys. Rev. D
14, 870--892 (1976)

\bibitem{Benjamin} Benjamin, S. C., Hayden, P. M.: quantum games and quantum
strategies. Phys. Rev. Lett. 87 06980 (2001)

\bibitem{Flitney4} Flitney, A. P., Hollenberg, L. C. L.: Nash equilibria in
quantum games with generalized two-parameter strategies. Phys. Lett. A 363
381-388 (2007)
\end{thebibliography}
\end{document}